\documentclass[10pt,letterpaper,twocolumn]{article}
\usepackage{ol2}
\usepackage[draft]{hyperref}
\usepackage{amsmath}
\begin{document}
\twocolumn[ 
\title{Dynamical electric dipole theory for quantitatively describing coupled split-ring resonators}
\author{Yong Zeng and Douglas H. Werner}
\address{Department of Electrical Engineering, Pennsylvania State
University, University Park, Pennsylvania 16802}
\begin{abstract}
Metallic split-ring resonators possess dominant electric dipoles
as well as considerable magnetic dipoles under proper excitations.
Full-wave numerical approaches are frequently employed to simulate
adjacent split-ring resonators, but simulations cannot explain the
underlying physics. An analytical theory based on a dynamic
electric dipole approximation is developed here. Detailed
theory-simulation comparisons demonstrate that this theory can
\textit{quantitatively} describe the interaction strength of
coupled split-ring-resonators under certain circumstances.
\end{abstract}
] 

Since its introduction in 1999, metallic split-ring resonators
(SRRs) have attracted intense attention because of their important
roles in artificial metamaterials \cite{john}. Under appropriate
illumination, circulating current will be excited inside the SRR
and lead to magnetic dipoles comparable in magnitude to electric
dipoles, making metallic SRRs ideal building blocks to achieve
strong magnetic responses in the optical region. A single SRR
consequently is referred to as a magnetic photonic atom
\cite{solymar}. Recently, several experiments are devoted to
multipole coupling between SRRs, especially electric and magnetic
dipole interaction \cite{hesmer,feth,sersic,liu2}. For instance,
planar metallic SRR arrays are fabricated with different lattice
spacing along different directions, and both magnetic and electric
near-field dipole coupling are found to simultaneously influence
the spectral positions of the plasmonic resonances \cite{sersic}.
Another experiment studied isolated metallic SRR dimers with
different configurations. Changing the relative magnetic dipole
phase of the constituent particles identifies the strength of
magnetic dipole coupling \cite{feth}. Two methods are frequently
employed to interpret the experimental observations, full-wave
numerical simulations and quasistatic coupling models. Simulations
are able to reproduce the experimental measurements
quantitatively, but shed little light on the underlying physics
\cite{feth}. The quasistatic model, on the other hand, obtains
variables by fitting experimental data. Consequently it can not
make an \textit{a priori} prediction
\cite{sersic,liu2,rechberger}.

In this Letter we will show that \textit{a priori} predictions of
the electromagnetic interaction are feasible for specific SRRs
structures. More particularly, we carefully compare the dynamic
radiation emitted from the magnetic dipole of a single SRR with
that from its electric counterpart. It is found that at particular
``safe" spatial regions the electric dipole exclusively dominates
the radiation field. An individual SRR consequently can be
approximated by an isotropic electric dipole whose polarizability
is determined by the particle geometry as well as the
illumination. This approximation can be further extended to
multiple SRRs where each constituent particle sits in the safe
zone of other particles. Two dimers are studied to verify our
theory. By comparing the analytical results with the corresponding
full-wave numerical results, it is demonstrated that our theory
allow us to predict the coupling strength \textit{quantitatively}.

We start with the optical response of a gold SRR as shown in Fig.
1(a), which closely matches the experimental samples of Ref.
\cite{sersic}. It is excited by an $x$-polarized plane wave
propagating in the $z$ direction. The absolute extinction and
absorption cross section of the particle are shown in Fig. 1(b)
\cite{yong1}. A peak corresponding to the fundamental plasmonic
resonance was found at a wavelength of 1352 nm. Because of the
structural symmetry breaking, a $z$-component magnetic dipole
$m_{z}$ is generated together with a dominant $x$-component
electric dipole $d_{x}$. Numerical simulations further suggest
that the ratio $\beta=-m_{z}/d_{x}/c$ is around $0.29i$
\cite{yong1}. The magnetic dipole therefore has an amplitude
comparable to its electric counterpart. In addition, because the
extinction spectra are very sensitive to the illumination
polarization, both electric and magnetic dipoles are strongly
anisotropic \cite{sersic,yong1}.

\begin{figure}[t]
\centering
\includegraphics[width=0.6\textwidth]{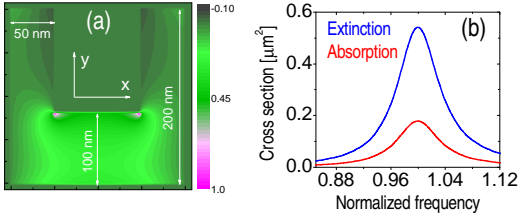}\vspace*{-11.0cm}
\caption{(a) The $x$-component of the current distribution of an
individual split-ring resonator at its fundamental resonance. The
incident pulse is $x$-polarized and propagating in the $z$
direction. (b) The extinction and absorption cross sections of the
single split-ring resonator. The frequency is normalized to the
fundamental resonant frequency $\omega_{0}$. The bulk plasma
frequency of gold is taken as $\omega_{p}=1.367\times 10^{16}
s^{-1}$, the phenomenological collision frequency
$\gamma=6.478\times 10^{13} s^{-1}$.} \label{fig1}
\end{figure}

It is well known that the electric field, in the near and far
zone, for an electric dipole source $\mathbf{p}$ is given by
\cite{jackson}
\begin{equation}
\mathbf{E}_{\textrm{e}}(\mathbf{r})=\left[A(kr)\mathbf{\underline{I}}+B(kr)\mathbf{\underline{T}}\right]\cdot\mathbf{p}
\label{eq1}
\end{equation}
with $k$ being the wave number, $\mathbf{\underline{I}}$ being an
unit tensor of rank 2. The tensor $\mathbf{\underline{T}}$ has
component as $T_{ij}=r_{i}r_{j}/r^{2}$, and the coefficients are
written as \cite{markel}
\begin{eqnarray}
&&A(kr)=\frac{e^{ikr}}{4\pi\epsilon_{0}r^{3}}\left(k^{2}r^{2}+ikr-1\right),\cr
&&B(kr)=\frac{e^{ikr}}{4\pi\epsilon_{0}r^{3}}\left(3-3ikr-k^{2}r^{2}\right).
\label{eq2}
\end{eqnarray}
The electric fields emitted from the electric dipole $d_{x}$ of a
single SRR are therefore direction-dependent: The $A$ term
survives for any radiation direction while the $B$ term vanishes
when $\mathbf{r}$ parallels the $y$ axis. In a similar way, the
radiation for a magnetic dipole source $\mathbf{m}$ bears the form
\begin{equation}
\mathbf{E}_{\textrm{m}}(\mathbf{r})=-\frac{Z_{0}k^{2}}{4\pi
r^{2}}e^{ikr}(\mathbf{r}\times\mathbf{m})(1+\frac{i}{kr}).
\label{eq3}
\end{equation}
with $Z_{0}=\sqrt{\mu_{0}/\epsilon_{0}}$ being the vacuum
impedance. The polarization of the radiation field therefore is
perpendicular to the plane defined by $\mathbf{r}$ and
$\mathbf{m}$. Consequently, the electric field emitted from the
magnetic dipole $m_{z}$ is polarized in the $xy$ plane and is
perpendicular to $\mathbf{r}$ when $\mathbf{r}$ lies in the $xy$
plane.

We now can analyze the radiation field of the single SRR in the
$xy$ plane. Because its fundamental wavelength, 1352 nm, is much
larger than its characteristic size, higher-order multipoles fall
off rapidly and the radiation emitted from the structure will come
mainly from its electric and magnetic dipoles when $kr>1$
\cite{jackson}. More specifically, when $\mathbf{r}$ parallels
$\mathbf{e}_{x}$, $\mathbf{E}_{\textrm{e}}$ is $x$-polarized and
$\mathbf{E}_{\textrm{m}}$ is $y$-polarized, and the ratio of their
amplitude $E_{\textrm{e}}/E_{\textrm{m}}$ equals $2/|\beta|kr$.
The magnetic dipole then dominates the radiation zone where
$kr\gg1$, and the electric dipole dominates a limited region where
$1<kr<1/|\beta|$. Similarly, the polarization of both
$\mathbf{E}_{\textrm{e}}$ and $\mathbf{E}_{\textrm{m}}$ are along
the $x$ axis for $\mathbf{r}$ parallel to $\mathbf{e}_{y}$. The
amplitude ratio $E_{\textrm{e}}/E_{\textrm{t}}$, with
$\mathbf{E}_{\textrm{t}}$ being
$\mathbf{E}_{\textrm{e}}+\mathbf{E}_{\textrm{m}}$, is further
given by
$\left[kr+i(1-k^{2}r^{2})\right]/\left[kr+|\beta|+i(1-k^{2}r^{2})\right]$.
As a result, we can safely neglect the magnetic dipole and use the
dynamic electric dipole approximation when $kr\gg|\beta|$. Similar
conclusions apply for $\mathbf{r}$ along the $z$ axis. The
electric vector $\mathbf{E}_{\textrm{m}}$ is exactly zero because
$\mathbf{r}$ parallels $m_{z}$, and the $\mathbf{E}_{\textrm{e}}$
is $x$-polarized. For a considerably large $kr$, the magnetic
dipole has no influence on the radiation field.

The discussions above can be extended to multiple SRRs, as long as
each constituent particle sits in the public ``safe" zone. To
validate the theory, we investigate two different configurations
of SRR dimers, side-by-side and on-top, which have been studied
experimentally \cite{feth}. The dipole polarizability
$\alpha(\omega)$ of an individual SRR is determined first, which
connects to the extinction cross-section spectrum $\sigma(\omega)$
as \cite{yong1}
\begin{equation}
\frac{Z_{0}}{|E_{0}|^{2}}\textrm{Re}\left[\mathbf{E}_{0}^{\ast}\cdot\int_{v}\mathbf{J}(\mathbf{r}')e^{-i\mathbf{k}_{0}\cdot\mathbf{r}'}d\mathbf{r}'\right]\approx
Z_{0}\omega\:\textrm{Im}\left[\alpha(\omega)\right], \label{eq4}
\end{equation}
where $\mathbf{E_{0}}$ stands for the incident electric field,
$\mathbf{J}$ represents the polarization current and the
integration is performed over the particle volume. We neglect the
retarded effect because the particle height, 25 nm, is much
shorter than the incident wavelength. Moreover, a Lorentz model
describes the polarizability $\alpha(\omega)$ around the resonance
\begin{equation}
\alpha(\omega)=\frac{4\pi\epsilon_{0}f}{\omega_{0}^{2}-\omega^{2}-i\omega\tau},
\label{eq5}
\end{equation}
where $\omega_{0}$ is the resonant frequency, $f$ stands for the
oscillator strength and $\tau$ measures the phenomenological
damping force \cite{jackson}. For the SRR shown in Fig. 1(a), we
find that $\tau/\omega_{0}\approx 0.072$ as well as $f/\tau\approx
1.29\times 10^{-5}$. It should be emphasized that the extinction
spectrum is experimentally measurable \cite{feth,husnik} and the
parameters of the Lorentz model can then be obtained by fitting
the experimental results.

By approximating each particle of an SRR dimer as an electric
dipole, a set of coupled equations describes its optical
properties
\begin{equation}
\mathbf{p}_{1}=\alpha(\omega)\left[\mathbf{E}_{0}+\mathbf{E}_{\textrm{e}}(\mathbf{d},\mathbf{p}_{2})\right],\:\:\:\
\mathbf{p}_{2}=\alpha(\omega)\left[\mathbf{E}_{0}+\mathbf{E}_{\textrm{e}}(\mathbf{d},\mathbf{p}_{1})\right],
\label{eq6}
\end{equation}
where $\mathbf{d}$ stands for the distance between
$\mathbf{p}_{1}$ and $\mathbf{p}_{2}$, with $\mathbf{p}_{1}$ and
$\mathbf{p}_{2}$ being the dipole moment of each particle,
respectively. The symmetry of the equations immediately implies
that the incident plane wave $\mathbf{E}_{0}$ only excites the
symmetrical (in-phase) mode in either configuration. The
extinction spectrum can be obtained by solving the set of
equations. For the side-by-side structure it is given by
\begin{equation}
\sigma_{e}(\omega,d)=\textrm{Im}\left[\frac{2Z_{0}\omega}{\alpha(\omega)^{-1}-A(kd)-B(kd)}\right],
\label{eq7}
\end{equation}
and for the on-top arrangement it is
\begin{equation}
\sigma_{e}(\omega,d)=\textrm{Im}\left[\frac{2Z_{0}\omega}{\alpha(\omega)^{-1}-A(kd)}\right].
\label{eq8}
\end{equation}
Because $A$ and $A+B$ tend to zero when $d$ goes to infinity, the
equations above lead to an intuitive fact: The total power taken
from the incident wave by two identical particles without coupling
is twice of that by a single particle. Furthermore, the resonant
frequency $\omega_{r}$ corresponds to the vanishing of the real
part of the denominator of $\sigma_{e}$ \cite{markel}, we
therefore have
\begin{equation}
4\pi\epsilon_{0}\:\textrm{Re}[A(\omega_{r}d/c)]=\frac{\omega_{0}^{2}-\omega_{r}^{2}}{f},
\label{eq9}
\end{equation}
for the on-top configuration, and
\begin{equation}
4\pi\epsilon_{0}\:\textrm{Re}[A(\omega_{r}d/c)+B(\omega_{r}d/c)]=\frac{\omega_{0}^{2}-\omega_{r}^{2}}{f},
\label{eq10}
\end{equation}
for the side-by-side configuration. Notice that the damping force
$\tau$ of the Lorentz model does not appear since it mainly
influences the bandwidth of the resonance \cite{jackson}. These
equations can be solved easily in the near-field zone where
$kr\ll1$. We find that $4\pi\epsilon_{0}\textrm{Re}[A]\approx
-1/d^{3}$ as well as $4\pi\epsilon_{0}\textrm{Re}[A+B]\approx
2/d^{3}$, the resultant $\omega_{r}$ is therefore bigger than
$\omega_{0}$ for the on-top dimer while smaller than $\omega_{0}$
for the side-by-side arrangement. In addition, the resonance shift
$|\omega_{r}-\omega_{0}|$ of the side-by-side structure is
considerably bigger than that of the on-top configuration
\cite{su}. The former structure therefore dominates the nearest
neighbor coupling of equally oriented SRRs in dense square arrays
\cite{sersic}. We want to stress that all the predictions here are
qualitatively consistent with the experimental observations
\cite{feth,sersic}.

\begin{figure}[t]
\centering
\includegraphics[width=0.6\textwidth]{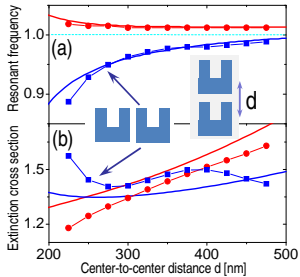}\vspace*{-10.5cm}
\caption{Two different configurations, side-by-side and on-top, of
two split-ring resonators. The resonant frequencies are normalized
to $\omega_{0}$, the fundamental resonant frequency of the
individual split-ring resonator. The extinction cross sections are
normalized to the $\sigma_{s}(\omega_{0})$, the extinction cross
section of the single split-ring resonator at its fundamental
resonance. The analytical and numerical results are shown with
solid and dotted curves, respectively.} \label{fig2}
\end{figure}

Equations (\ref{eq9}) and (\ref{eq10}) are solved for the
separation $d$ ranging from 200 to 500 nm, and the results are
plotted in Fig. 2(a) with solid curves. The resonant frequencies
$\omega_{r}$ are normalized to $\omega_{0}$ of the individual
particle. A finite-difference time-domain method is then applied
to numerically simulate the two dimers, and the calculated results
are plotted in Fig. 2(a) with dotted curves \cite{taflove}. The
center-to-center distance $d$, as marked in Fig. 2, is gradually
increased from 225 nm to 475 nm (corresponding to $kd$ from 1.0 to
2.2) with an increment of 25 nm. Notice that the width of the gap
between these two particles is given by $d-200$ nm; they are very
close to each other. The numerical result of the on-top dimer is
found to agree perfectly with its analytical counterpart when
$d\geq 250$ nm, while a slight discrepancy appears for the
side-by-side configuration: The numerical result oscillates around
its analytical counterpart with an average relative difference
smaller than 1\%. It is possibly induced by the non-uniform
distribution of the polarization current inside the SRR. As shown
in Fig. 1(a), the current is found to be strongly localized around
the two inner corners. Assuming the current ``hot" points are the
positions of the electric dipoles, the dipole-to-dipole distance
is shortened to roughly $d-100$ nm for the side-by-side dimer.
Higher-order multipoles such as an electric quadrupole may
interfere with the electric dipole to a certain degree.

Although our theory describes the coupling strength
\textit{quantitatively}, it can not be extended to the extinction,
absorption or scattering spectra because these quantities need the
information regarding radiation along all the directions. However,
its predictions can serve as a first approximation since the
dominant electric dipole is contained. Take the two SRR dimers
studied above as examples: their normalized extinction cross
sections versus $d$ are computed analytically and numerically, and
the results are plotted in Fig. 2(b). Unlike the resonant
frequency, the analytical results of the extinction are
considerably different from their numerical counterparts, with an
average relative difference around 10\%. In addition, the theory
must be employed cautiously to study the gap-to-gap and
back-to-back configurations considered in Ref. \cite{feth}. The
constituent particle in either dimer has in-phase electric dipoles
but out-of-phase magnetic dipoles; Equation (6) then loses its
validity. It is still a good approximation when $d$ is big enough
because the magnetic dipole coupling falls off rapidly with
increasing inter-particle spacing. Finally, note that
magnetoinductive coupling will appear in the side-by-side dimer
when $d\gg c/|\beta|\omega_{0}$, as suggested by our theory
\cite{solymar}.

To summarize, by carefully analyzing the fields emitted from the
electric dipole and magnetic dipole of an individual split-ring
resonator, we demonstrate that in certain circumstances a simple
dynamic electric dipole can be employed to approximate the
split-ring resonator. Two configurations of coupled split-ring
resonators are studied to validate our theory by detailed
theory-simulation comparisons. It is shown that this theory can
predict coupling strength quantitatively and other optical
quantities such as extinction cross section qualitatively.

We thank Lujun Huang for assistants. This work was supported in
part by the Penn State Materials Research Science and Engineering
Center under National Science Foundation (NSF) grant no. DMR
0213623.


\begin{thebibliography}{99}
\bibitem{john} J. B. Pendry, A. J. Holden, D. J. Robbins, and W. J.
Stewart, IEEE Trans. Microwave Theory Tech. 47, 2075 (1999); D. R.
Smith, W. J. Padilla, D. C. Vier, S. C. Nemat-Nasser, and S.
Schultz, Phys. Rev. Lett. 84, 4184 (2000); S. Linden, C. Enkrich,
M. Wegener, J. Zhou, T. Koschny, and C. M. Soukoulis, Science 306,
1351 (2004); R. Merlin, Proc. Natl. Acad. Sci. U.S.A. 106, 1693
(2009); M. Decker, S. Linden, and M. Wegener, Opt. Lett. 34, 1579
(2009); A. E. Nikolaenko, F. De Angelis, S. A. Boden, N.
Papasimakis, P. Ashburn, E. Di Fabrizio, and N. I. Zheludev, Phys.
Rev. Lett. 104, 153902 (2010).
\bibitem{solymar} L. Solymar and E. Shamonina, {\it Waves in Metamaterials} (Oxford University, 2009).
\bibitem{hesmer} F. Hesmer, E. Tatartschuk, O.
Zhuromskyy, A. A. Radkovskaya, M. Shamonin, T. Hao, C. J. Stevens,
G. Faulkner, D. J. Edwards, and E. Shamonina, Phys. Status Solidi
B 244, 1170 (2007); N. Liu, S. Kaiser, and H. Giessen, Adv. Mater.
20, 4521 (2008); M. Decker, S. Burger, S. Linden, and M. Wegener,
Phys. Rev. B 80, 193102 (2009).
\bibitem{feth} N. Feth, M. K\"{o}nig, M. Husnik, K. Stannigel, J. Niegemann, K. Busch, M. Wegener, and S.
Linden, Opt. Express 18, 6545 (2010).
\bibitem{sersic} I. Sersic, M. Frimmer, E. Verhagen, and A. F.
Koenderink, Phys. Rev. Lett. 103, 213902 (2009).
\bibitem{liu2} N. Liu, H. Liu, S. Zhu, and H. Giessen, Nat. Photonics 3, 157 (2009).
\bibitem{rechberger} W. Rechberger, A. Hohenau, A. Leitner, J. R. Krenn, B. Lamprecht,
and F. R. Aussenegg, Opt. Commun. 220, 137 (2003); H. Liu, D. A.
Genov, D. M. Wu, Y. M. Liu, J. M. Steele, C. Sun, S. N. Zhu, and
X. Zhang, Phys. Rev. Lett. 97, 243902 (2006).
\bibitem{yong1} Y. Zeng, C. Dineen, and J. V. Moloney, Phys. Rev. B 81, 075116 (2010).
\bibitem{jackson} J. D. Jackson, {\it Classical Electrodynamics}, 3rd ed. (Wiley, New York, 1999).
\bibitem{markel} V. A. Markel, J. Opt. Soc. Am. B 12, 1783 (1995).
\bibitem{husnik} M. Husnik, M. W. Klein, N. Feth, M. K\"{o}nig, J. Niegemann, K. Busch, S. Linden, and M.
Wegener, Nat. Photonics 2, 614 (2008).
\bibitem{su} K. H. Su, Q. H. Wie, X. Zhang, J. J. Mock, D. R. Smith, and S.
Schultz, Nano Lett. 3, 1087 (2003); P. Olk, J. Renger, M. T.
Wenzel, and L. M. Eng, Nano Lett. 8, 1174 (2008); A. M. Funston,
C. Novo, T. J. Davis, and P. Mulvaney, Nano Lett. 9, 1651 (2009);
J. Petschulat, C. Menzel, A. Chipouline, C. Rockstuhl, A.
T\"{u}nnermann, F. Lederer, and T. Pertsch, Phys. Rev. A 78,
043811 (2008); N. A. Gippius, T. Weiss, S. G. Tikhodeev, and H.
Giessen, Opt. Express 18, 7569 (2010).
\bibitem{taflove} A. Taflove and S. C. Hagness, {\it Computational Electrodynamics: the finite-difference time-domain method}, 2nd ed. (Artech
House, Boston, 2000); Y. Zeng and J. V. Moloney, Opt. Lett. 34,
1600 (2009).
\end{thebibliography}
\end{document}